\documentclass[fleqn,usenatbib]{mnras}
\usepackage{newtxtext,newtxmath}
\usepackage[T1]{fontenc}
\DeclareRobustCommand{\VAN}[3]{#2}
\let\VANthebibliography\thebibliography
\def\thebibliography{\DeclareRobustCommand{\VAN}[3]{##3}\VANthebibliography}

\usepackage{graphicx}	
\usepackage{amsmath}	

\usepackage{amssymb}	

\usepackage{soul}

\defcitealias{Gereb2015}{G15}
\newcommand{\ie}{{\emph{i.e. }}}
\newcommand{\eg}{{\emph{e.g. }}}
\newcommand{\kmps}{$\,\rm{km} \, \rm{s}^{-1}$\,}
\newcommand{\sqdeg}{$\,{\rm{deg}}^{2}$ }
\newcommand{\sqarcsec}{$\,{\rm{arcsec}}^{2}$ }
\newcommand{\sqarcmin}{$\,{\rm{arcmin}}^{2}$ }

\title[HI absorption in Norma's BCG]{H{\sc i} absorption associated with Norma's brightest cluster galaxy}

\author[M. Saraf et al.]{
Manasvee Saraf,$^{1,2,3}$\thanks{E-mail: manasvee.saraf@icrar.org}
O. Ivy Wong,$^{2,1}$
Luca Cortese$^{1,3}$
and B\"{a}rbel S. Koribalski$^{4,5}$
\\
$^{1}$International Centre for Radio Astronomy Research, The University of Western Australia, Crawley, WA 6009, Australia\\
$^{2}$Australia Telescope National Facility, CSIRO, Space and Astronomy, P.O. Box 1130, Bentley, WA 6102, Australia\\
$^{3}$ARC Centre of Excellence for All-Sky Astrophysics in 3 Dimensions (ASTRO 3D), Australia\\
$^{4}$Australia Telescope National Facility, CSIRO, Space and Astronomy, P.O. Box 76, Epping, NSW 1710, Australia\\
$^{5}$School of Science, Western Sydney University, Locked Bag 1797, Penrith, NSW 2751, Australia
}
\date{Accepted 2022 December 12. Received 2022 December 08; in original form 2022 May 13}

\pubyear{2022}

\begin{document}
\label{firstpage}
\pagerange{\pageref{firstpage}--\pageref{lastpage}}
\maketitle

\begin{abstract}
ESO 137-G006 is the brightest cluster galaxy (BCG) of the cool-core and dynamically young Norma cluster. 
We discover an atomic hydrogen (H{\sc i}) absorption line associated with this BCG using the Australia Telescope Compact Array.
We estimate a gas column density of $ \approx (1.3 \pm 0.2) \times 10^{20}\,T_{\rm{spin}}$ atoms cm$^{-2}$ with spin temperature, $T_{\rm{spin}} \leq 194$ K, consistent with the H{\sc i} properties of other early-type galaxies and cool-core cluster BCGs.
The relationship between the presence of cold gas and a cluster cooling flow is unclear. 
Our results support the scenario that ESO 137-G006 may be a recent arrival to the cluster centre and not the original BCG. 
This scenario is consistent with the observed spatial alignment of the BCG’s wide-angle tail radio lobes with Norma's X-ray sub-cluster and the significant line-of-sight velocity offset between the mean velocity of Norma and that of the BCG.
\end{abstract}



\section{Introduction}
\label{sec:Start}
A brightest cluster galaxy (BCG) is either the most luminous or the most massive galaxy residing at the bottom of its host cluster's potential well. 
About 20\% of BCGs are cD type galaxies, \ie supergiant ellipticals \citep{Seigar2007, Coziol2009}.
cD galaxies have diffuse envelopes of old stellar population that are $\sim 10$ times larger than that of a typical elliptical.
cD BCGs are theorised to have grown via minor mergers and tidal stripping events linked to cluster formation \citep{Dubinski1998, Coziol2009, DePropris2021}.
Although the gas reservoirs of cD galaxies are generally depleted, cold gas reservoirs have been observed in several cD BCGs at the centre of cool-core clusters \citep[][and references therein]{Hogan2014, Morganti2018}.

Cool-core clusters exhibit strongly peaked X-ray profiles and short central cooling times \citep[$ t_{\rm{cool}}< 7$ Gyr,][]{Mittal2009}. 
Their central `cooling flow' is thought to be responsible for the presence of cold gas in their BCGs \citep[\eg][]{Hogan2014}. 
Here we present the case of a cool-core cluster cD BCG hosting cold gas of origin that cannot simply be attributed to the presence of a cooling flow. 

The Norma cluster (ACO 3627) is a nearby ($z = 0.016$, $d \approx 70$ Mpc) and massive ($M \sim 10^{15}\,h_{73}^{-1}\,\mathrm{M}_{\odot}$) X-ray bright ($2 \times 10 ^{-10}$ ergs s$^{-1}$ cm$^{-2}$) galaxy cluster \citep[][]{Boehringer1996, Woudt2008}. 
Norma is considered a younger analogue of the Coma cluster. 
Coma is a more virialised and distant but a similarly massive and X-ray bright cluster compared to Norma \citep{Kraan-Korteweg1996, Woudt1998, Bravo-Alfaro2000}. 
The Norma cluster, like the Coma cluster, hosts two bright cD galaxies.
The two brightest galaxies in the Norma cluster are ESO 137-G006 and ESO 137-G008, with the Two Micron All Sky Survey (2MASS) $K$-band luminosity of $L_{Ks} = 10^{11.76} \, \mathrm{L}_{\odot}$ \citep{Sun2007, Sun2009}.
ESO 137-G006 and ESO 137-G008 are offset by 5.5 arcmin ($\sim 109$ kpc) and 10.5 arcmin ($\sim 208$ kpc) respectively from the cluster core, defined by the peak X-ray emission \citep{Boehringer1996, Sun2007}.
Following \citet{Sun2009}, we identify ESO 137-G006 to be the main BCG of the Norma cluster as it is located closer to the peak of the X-ray emission and has the brightest 1.4 GHz luminosity in the cluster \citep[$L_{1.4 \, \rm{GHz}} = 10^{25.38} \, \rm{W \, Hz^{-1}}$,][]{Sun2007}.

A sub-cluster, that is likely to be merging into the main cluster, has been observed in the X-ray distribution of Norma \citep{Boehringer1996}.  
In combination with the observed spatial offset between the cD galaxies and the cluster core, this ongoing sub-cluster merger is consistent with Norma being a dynamically-young cluster \citep{Coziol2009, Hamer2012, Hashimoto2014, DePropris2021}. 

ESO 137-G006 (WKK 6269) is the focus of this study and has an X-ray cool-core of the luminous corona type associated with its active galactic nucleus \citep[AGN,][]{Sun2007, Sun2009, Nishino2012}. 
This BCG is one of two radio galaxies near Norma's core and has strong radio lobes \citep[$L_{1.4 \, \rm{GHz}} = 10^{25.38} \, \rm{W \, Hz^{-1}}$,][]{Sun2007} bent in a `C' shaped wide-angle tail (WAT) morphology \citep{Jones1996}. 
The presence of cold gas is understood to be key in the formation of such strong radio lobes \citep[\eg][]{Bicknell1995, Morganti2005, Emonts2008}. 
The WAT is spatially aligned (in projection) with the X-ray sub-cluster, placing this BCG at the leading edge of the sub-cluster merger. 
The observed line-of-sight velocity offset between the mean cluster velocity ($4871 \pm 54$ \kmps) and that of ESO 137-G006 ($5441 \pm 52$ \kmps) provides further support to the sub-cluster merger hypothesis \citep{Woudt2008}. 
Despite being a massive cluster currently in formation, Norma is not well-studied due to its location behind the Galactic plane \citep{Jones1996, Woudt1998}. 

Radio emission is not attenuated by the Galaxy in the foreground. 
The 21-cm line traces the spin-flip transition of atomic hydrogen (H{\sc i}). 
Thus, observations at radio wavelengths can enable us to probe the H{\sc i} content of ESO 137-G006. 
With the current technological advancements, dense H{\sc i} columns ($N_{\rm{HI}} \geq 10^{19}$ cm$^{-2}$) are detected in absorption against radio-loud AGN \citep[$L_{1.4 \, \rm{GHz}} \geq 10^{24} \, \rm{W \, Hz^{-1}}$,][]{Sun2009, Allison2022}. 
The absorption profiles allow the investigation of cold gas properties such as the H{\sc i} distribution, dynamics, and kinematics \citep[\eg][]{Morganti2001, Morganti2006, Oosterloo2010, Curran2010, Chandola2013, Allison2015, Allison2016, Moss2017, Grasha2019, Deb2020, Mahony2021, Kerrison2021}. 
Moreover, radio interferometers are able to spatially resolve the core of ESO 137-G006 due to Norma's proximity \citep[\eg][]{Ramatsoku2020}. 

In this study, we use Australia Telescope Compact Array (ATCA) observations to conduct a search for H{\sc i} towards the core region of ESO 137-G006 with higher sensitivity and resolution than previous studies \citep[\eg][]{Allison2014}. 
Our observation suggests that cold gas is associated (within 5 kpc) with this BCG of a very massive but dynamically young cool-core cluster.
Thus, we find that some BCGs may retain or attain cold gas in the process of being accreted into the cluster's core during later stages of cluster formation.

This paper describes our search of H{\sc i} absorption towards the core of Norma's BCG, ESO 137-G006. 
Section \ref{sec:Obs} describes the ATCA observations of high spectral resolution and our adopted data reduction process to optimise these observations for high spatial resolution and sensitivity.
Section \ref{sec:Results} presents the resulting H{\sc i} absorption detection associated with ESO 137-G006 and the analysis we conducted on the location, kinematics, density and temperature of the H{\sc i}. 
Section \ref{sec:Discussion} discusses the cold gas and stellar properties of ESO 137-G006 and the implications of this H{\sc i} detection on our understanding of cold gas in massive ($M \sim 10^{15}\,\mathrm{M}_{\odot}$) cool-core cluster BCGs. 
The limitations of our work and the possible advancements that can be made with upcoming radio surveys and new high resolution multi-wavelength data are also discussed in Section \ref{sec:Discussion}. 
We summarise our findings in Section \ref{sec:End}. 

Throughout this paper we assume a $\Lambda$ cold dark matter cosmology with $\Omega_{M} = 0.3$ and $\Omega_{\Lambda} = 0.7$, and a Hubble constant of $H_{0} = 70$ \kmps Mpc$^{-1}$.

\section{Observations and Data Reduction}
\label{sec:Obs}
\subsection{ATCA Observations}

We acquire the ATCA observations of Norma's central $\approx 1$ \sqdeg from project C2891, conducted in October 2013. 
We use the observations made with the 6A array configuration centred at 1396 MHz. 
The configuration of the Compact Array Broadband Backend (CABB) is the 64-MHz zoom mode \citep[with 64-MHz bandwidth across 2049 32-kHz channels,][]{Wilson2011}. 
From this CABB configuration, we obtain a spectral resolution of 6.6 \kmps in radio velocity. 
Observations from the 6A array configuration, with baselines ranging from 337m to 5939m, can be optimised for the high spatial resolution required to resolve the core of ESO 137-G006. 

The observations were taken in mosaicking mode, which consisted of 26 pointings. 
These pointings were observed for $\approx 2$ min each, followed by an $\approx 4.3$ min scan of the nearby phase calibrator, PKS 1613-586, to enable time-varying bandpass and gain corrections. 
This cycle was repeated 10-12 times for improved $uv$-coverage. 
The ATCA primary calibrator, PKS 1934-638, was observed for $\approx 7$ min to enable the calibration of the absolute flux scale and the bandpass (see appendix \ref{sec:ObsTable} for more details). 

We integrate over $\approx 11$h of observations to achieve optimal $uv$-coverage. 
Since these observations were conducted at night, they have minimal effects of solar interference. 
We only reduce data from pointings 12, 13, 14, 17, 18, 19 and 23 because ESO 137-G006 is located in the intersecting region between these pointings (see Fig. \ref{fig:Maps}a). 

\subsection{Data Reduction}

The observations are processed using the \texttt{MIRIAD} software \citep{Sault1995}. 
We follow the standard procedures for continuum and spectral-line data reduction to optimise for high spatial resolution and sensitivity. 

We flag, calibrate and continuum-subtract the $uv$-data in the following manner. 
Using \texttt{ATLOD}, we load the $uv$-data at the H{\sc i} rest frequency (1.4204058 GHz) and flag channels at the edge and with known radio-frequency interference (RFI). 
We flag the remaining RFI emission using \texttt{UVFLAG}, \texttt{BLFLAG} and \texttt{PGFLAG}. 
Using \texttt{UVFLUX}, we compare the theoretical root mean square (RMS) noise of the observations with the RMS noise of the flagged $uv$-data to ensure that the sensitivity of our flagged dataset has not been significantly diminished by the flagging process. 
We confirm that the RMS noise of the flagged $uv$-data for all our pointings are comparable with their respective theoretical RMS noise and that there are no unexpected peaks in the mean amplitude of each pointing’s $uv$-data.
Next, we calibrate the $uv$-data with the primary calibrator, using \texttt{MFCAL}, and with the phase calibrator, using \texttt{GPCAL} and \texttt{GPCOPY}. 
Due to remaining RFI, we flag all data within the velocity range, 7668 \kmps $< v <$ 9056 \kmps, using \texttt{UVFLAG}.

We separate each of the pointing's continuum and line $uv$-data individually using \texttt{UVLIN}. 
\texttt{UVLIN} models the continuum in all the line-free channels as a polynomial to perform continuum subtraction or extraction.
We fit a low order polynomial to represent the continuum spectrum for each pointing's $uv$-data (see appendix \ref{sec:ContSub} for more details).
We conduct the continuum subtraction in the visibility plane as done in similar previous studies \citep[\eg][]{Reeves2016, Maccagni2018}.
However, due to the strong and extended continuum emission within the target field, there remains a residual ripple in the baseline of the line spectrum after imaging.
We account for this ripple by conducting baseline subtraction as explained in section \ref{sec:Results}.

We image the continuum and line $uv$-data separately in the following manner. 
Using \texttt{INVERT}, we perform inverse Fourier transform on the pointings individually. 
We set \texttt{INVERT} to create `dirty' cubes, in Stokes I polarisation, with pixels of size 2 arcsec centred on ESO 137-G006 (R.A. (J2000) = 16:15:04 and Dec (J2000) = $-60$:54:26). 
To achieve an optimal balance between signal-to-noise ratio (SNR) and resolution, we weight the baselines with robustness $-0.5$ \citep[][]{Briggs1995}. 
As a result, the dirty cubes have a theoretical RMS noise, $\sigma \approx 9$ mJy beam$^{-1}$, with a synthesised Gaussian beam (yellow hatched ellipse in Fig. \ref{fig:Maps}c) of full width at half-maximum, FWHM = 9 arcsec $\times$ 7 arcsec, and position angle, PA = $31^{\circ}$ (measured counterclockwise from North throughout this paper).
To perform deconvolution, we produce Clean component models (with cut-off $= 5\sigma$) using \texttt{CLEAN} and restore the dirty cubes with these models using \texttt{RESTOR}. 
Finally, we use \texttt{LINMOS} to linearly mosaic the individual `clean' cubes of the 7 pointings. 
To optimise for dynamic range, we set \texttt{LINMOS} to perform primary beam correction, with tapering, at a cut-off of 0.4 primary beam response. 
We obtain two 64-MHz bandwidth clean mosaic cubes - a continuum cube and an H{\sc i} spectral cube with RMS noise of 155 mJy beam$^{-1}$ and 34 mJy beam$^{-1}$ respectively. 

Additionally, we create a total intensity continuum image at 1396 MHz (presented in Fig. \ref{fig:Maps}). 
We use \texttt{INVERT} to perform multi-frequency synthesis imaging on the unsubtracted $uv$-data. 
We set a $5\sigma$ cutoff in \texttt{CLEAN} using the theoretical RMS noise of the dirty image, 0.3 mJy beam$^{-1}$. 
The RMS noise in the resulting clean image is 6 mJy beam$^{-1}$. 

We use the Cube Analysis and Rendering Tool for Astronomy \citep[\texttt{CARTA,}][]{Comrie2021} to inspect the spectral cube and the total intensity continuum image.
The WAT morphology of the radio continuum core and lobes of ESO 137-G006 observed in our ATCA data is consistent with that reported by \citet[][]{Ramatsoku2020}.
In the spectral cube, we detect H{\sc i} in absorption against the radio core.
We detect no H{\sc i} in emission.

To constrain the extent of the H{\sc i} absorption, we first constrain the spatial extent of the background continuum source.
To find and fit the continuum source, we use the \texttt{IMFIT} task of \texttt{MIRIAD} on the total intensity continuum image, within a square aperture of length 22 arcsec centred on R.A. (J2000) = 16:15:04 and Dec (J2000) = $-60$:54:26. 
This square aperture maximises the coverage of the radio core, while excluding any structure from the WAT, and has a RMS residual of 5 mJy beam$^{-1}$.
\texttt{IMFIT} returns a 10 arcsec $\times$ 9 arcsec Gaussian at PA $57^{\circ}$ centred on R.A. (J2000) = 16:15:03.7 and Dec (J2000) = $-60$:54:24.9. 
This Gaussian has a total integrated continuum flux, at 1396 MHz, of 65 $\pm$ 18 mJy and is comparable to the size of the synthesised beam.
However, integrating across just one FWHM of the synthesised beam underestimates the integrated flux density of the H{\sc i} absorption feature by 1.2 Jy$\,$\kmps (\ie by $4\sigma$, see section \ref{sec:HIfit}).
Thus, we extract the spectrum by integrating across an elliptical aperture twice the FWHM of the synthesised beam (blue ellipse in Fig. \ref{fig:Maps}c).

We extract spectra of total flux density as a function of optical velocity using \texttt{CARTA}.
Spectra are extracted from an 18 arcsec $\times$ 14 arcsec elliptical aperture at PA $31^{\circ}$ centred on R.A. (J2000) = 16:15:03.7 and Dec (J2000) = $-60$:54:24.9.
We refer to this elliptical aperture as the H{\sc i} detection region throughout this paper.
We extract a line spectrum from the spectral cube and, to calculate the H{\sc i} optical depth, a continuum spectrum from the continuum cube (see section \ref{sec:NHI}).
To distinguish the edges of the H{\sc i} profile, we perform Hanning smoothing (with kernel = 3) on both the extracted spectra.
In section \ref{sec:Results}, we analyse and present the H{\sc i} properties of the smoothed line and continuum spectra. 
We present the spectra without smoothing in appendix \ref{sec:OGspectra} for completeness and reproducibility.
We find no significant differences between the H{\sc i} properties of the spectra with and without smoothing (see table \ref{tab:HIprop}).
We also extract the spectra from the aforementioned square aperture and perform the same analysis as in section \ref{sec:Results}, and find no significant differences between the H{\sc i} properties of the spectra from the elliptical aperture and those from the square aperture (see appendix \ref{sec:BoxSpectra} for details).

\section{Results and Analysis}
\label{sec:Results}
In this section, we present the total intensity continuum image at 1396 MHz of the ATCA observations towards ESO 137-G006 and the detected H{\sc i} absorption spectrum.
We calculate and analyse the H{\sc i} properties of this detection.
The uncertainties have been calculated as per standard error propagation unless otherwise specified. 

\subsection{Spatial Analysis}

\begin{figure}
    \centering
    \includegraphics[width=0.88\columnwidth]{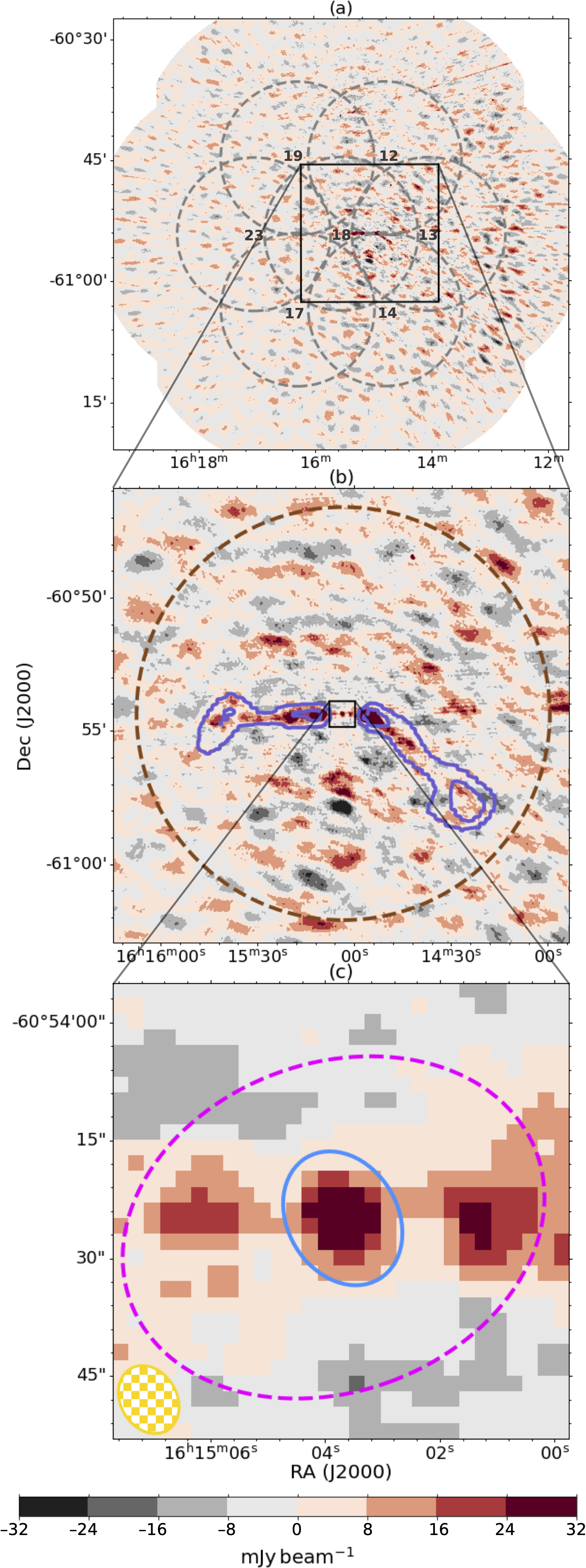}
    \caption{
    The reduced total intensity continuum image of the ATCA observations towards ESO 137-G006.
    The image has a RMS noise level of $8$ mJy beam$^{-1}$. 
    Zoom panel b shows the SUMSS structure of the BCG's WAT (purple contours) and the HIPASS synthesised beam (brown dashed circle) centred on the radio core. 
    Zoom panel c shows that we detect H{\sc i} against the radio core in an elliptical aperture (blue ellipse) twice the FWHM of the ATCA data's synthesised beam (yellow hatched ellipse).
    The H{\sc i} is constrained within the 2MASS isophotal diameter of ESO 137-G006's stellar component (magenta dashed ellipse).}
    \label{fig:Maps}
\end{figure}

\subsubsection{Detection location}

The total intensity continuum image at 1396 MHz is displayed in Fig. \ref{fig:Maps} with different zoom panels. 
Fig. \ref{fig:Maps}a displays the entire area of the reduced data. 
The dashed-grey circles (of radius 9.5 arcmin) are centred on the 7 ATCA pointings and mark their coverage with more than 0.5 primary beam response. 
The overlapping region between these dashed circles is a square of length 22.5 arcmin centred on R.A. (J2000) = 16:15:04 and Dec (J2000) = $-60$:54:26. 
Within this overlapping region, the RMS noise level in the spectral cube is 6 mJy beam$^{-1}$ and in the continuum total intensity image is $8$ mJy beam$^{-1}$.
The discrete colour scale used in all 3 panels of Fig. \ref{fig:Maps} corresponds to integer factors of the RMS noise level in the continuum image.
Shades of red and grey map the positive and negative intensities respectively. 

The 17 arcmin $\times$17 arcmin zoomed panel (Fig. \ref{fig:Maps}b) around ESO 137-G006, displays the radio continuum structure of the BCG. 
The Sydney University Molonglo Sky Survey's (SUMSS) 843 MHz data (at 3 and 5 times the RMS noise level of 0.9 Jy beam$^{-1}$) are overlaid as purple contours for comparison with our 1396 MHz observations.
The brown dashed circle (of radius 15.5 arcmin) in Fig. \ref{fig:Maps}b marks the H{\sc i} Parkes All Sky Survey (HIPASS) synthesised beam. 
In our reduced ATCA data, we obtain a more than 100 times smaller synthesised beam, marked as the yellow hatched ellipse in the 1 \sqarcmin zoomed panel (Fig. \ref{fig:Maps}c). 
Using this higher resolution data, we spatially locate the H{\sc i} source between the host BCG's WAT, centred on the unresolved radio continuum source.
This H{\sc i} detection region is marked by the blue ellipse in Fig. \ref{fig:Maps}c.
The H{\sc i} source spatially lies within ESO 137-G006's diameter, defined by the 2MASS 20.0 $K$-mag arcsec$^{-2}$ isophotal diameter of 55 arcsec $\times$ 41.8 arcsec at PA $-70^{\circ}$ (magenta dashed ellipse in Fig. \ref{fig:Maps}c). 
We do not detect H{\sc i} in emission or absorption, across a minimum of 2 channels and one beam, anywhere else in the spectral cube including against the radio lobes of ESO 137-G006. 
Given the spectral resolution and the noise level in the spectral cube, our upper limit on the integrated spectral line is 0.2 Jy$\,$\kmps.

\subsubsection{Compact radio source}

In the 1396 MHz ATCA continuum observations, we find an unresolved source of total integrated continuum flux, 65 $\pm$ 18 mJy, at the coordinates of ESO 137-G006 (Fig.\ref{fig:Maps}c). 
This unresolved continuum source is consistent in location, size and flux density with the 167 mJy point source detected by \citet{Ramatsoku2020} using the 1398 MHz Karoo Array Telescope (MeerKAT) data of higher resolution (7.5 arcsec synthesised beam) and sensitivity (20.8 $\mu$Jy beam$^{-1}$ RMS noise). 
At the distance of ESO 137-G006, 1 arcsec is equivalent to 0.33 kpc \citep[see][]{Ramatsoku2020}. 
Thus, in our ATCA observations of synthesised beam 9 arcsec $\times$ 7 arcsec, an unresolved source has a size upper limit of < 3 kpc.
According to the definition used in \citet{ODea2021}, compact steep-spectrum (CSS) sources have typical sizes < 20 kpc and  gigahertz peaked-spectrum (GPS) sources have typical sizes < 0.5 kpc.
The size upper limit of < 3 kpc on the detected unresolved continuum source and the radio lobes' morphology on both sides of this source suggest that it may be a compact symmetric object \citep[CSO;][]{ODea2021}.
A greater fraction of compact radio sources than extended radio sources are detected with H{\sc i} absorption \citep{Morganti2001, Gupta2006, Chandola2013, Gereb2014, Grasha2019, ODea2021}.
Hence, our H{\sc i} absorption detection is likely to be against a compact continuum core coincident with the centre of ESO 137-G006.

We can get more information about the evolutionary stage of ESO 137-G006's continuum core from its radio spectral index, $\alpha$, defined by: 
\begin{equation}
    S_{\nu} \propto \nu^{\alpha}
    \label{eq:SI}
\end{equation}
where $S_{\nu}$ is the flux density at frequency $\nu$. 
In addition to our ATCA observation at 1.4 GHz, we find a previous ATCA observation at 18 GHz \citep{Burke-Spolaor2009} with a total continuum flux density of 140 $\pm$ 50 mJy towards the core of ESO 137-G006. 
Using the two flux density measurements at 1.4 GHz and 18 GHz, we estimate the core's spectral index to be $\alpha = 0.3 \pm 0.2$. 
This spectral index, \ie $\alpha > -0.5$, is consistent with that found by \citet{Ramatsoku2020} and suggests that ESO 137-G006's  continuum core is a flat-spectrum source \citep[\eg Fig. 8 in][]{Sadler2016}.
On the other hand, many more GPS sources, than flat-spectrum sources, have been detected with absorption in early-type galaxies \citep[ETGs, \eg][]{Morganti2001, Chandola2013, Gereb2014, Glowacki2017, Grasha2019}. 
So, we think future continuum measurements of ESO 137-G006's radio core at other frequencies (between the range 1.4-18 GHz) could reveal a peak in the spectrum, suggesting that the core is a GPS source \citep[\eg][]{Gupta2006, Chandola2013, Kerrison2021}. 
Nevertheless, the spectral index of ESO 137-G006's core suggests that we may have detected H{\sc i} absorption against a young compact radio source with restarting AGN activity.

\subsection{Spectral Analysis}
\label{sec:HIfit}

\begin{figure*}
    \centering
    \includegraphics[width=\textwidth]{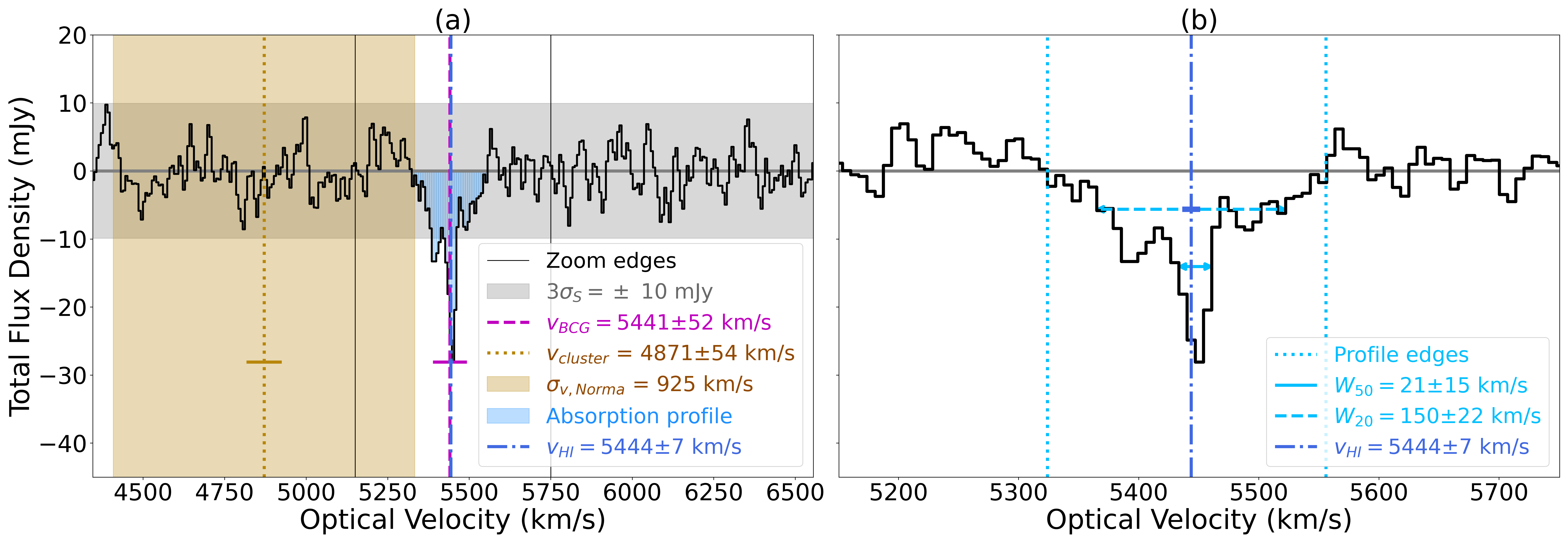}
    \caption{
    The total flux density spectrum integrated over the H{\sc i} detection region of the elliptical aperture towards ESO 137-G006's core.
    The velocity centroid of the H{\sc i} absorption profile is marked as the blue dash-dot line.
    In panel a, the velocities of the BCG (magenta dashed) and Norma (golden dotted) are marked as vertical lines with error-bars.
    The blue, horizontal grey and vertical golden shaded regions mark the H{\sc i} profile area, the 3 times RMS spectral noise and the velocity dispersion of the Norma cluster respectively.
    Zoom-panel b marks the edges (vertical cyan dotted lines), $W_{50}$ (horizontal cyan solid line), $W_{20}$ (horizontal cyan dashed line) of the absorption profile.
    Together with Fig. \ref{fig:Maps}c, this figure shows that the detected H{\sc i} absorption line is statistically significant and associated with Norma's BCG.}
    \label{fig:HIspec}
\end{figure*}

The line spectrum is extracted from the H{\sc i} detection region in the spectral cube and smoothed by a Hanning kernel of 3.
This spectrum is subtracted by a second order polynomial, fit within 3940 \kmps $<v< 6961$ \kmps, to remove the residual baseline ripple in the velocity range around the H{\sc i} absorption profile (see Fig. \ref{fig:OGspec} for the full bandwidth perspective and appendix \ref{sec:OGspectra} for details of the baseline subtraction).
The total bandwidth of the observation is $-$1610 \kmps $< v < $ 12381 \kmps, but only a fraction of this total bandwidth around the peak of the H{\sc i} profile is displayed in the panels of Fig. \ref{fig:HIspec}.

Zoom panel Fig. \ref{fig:HIspec}b displays the velocity range, 5150 \kmps $< v < $ 5751 \kmps, and is used to fit the H{\sc i} profile. 
The peak of the profile is observed at $v_{\rm{peak}} = 5451 \pm 7$ \kmps.
We do not fit a functional form to the absorption line due to the narrow width of the profile relative to the spectral resolution of our observations.
Instead, we define the profile edges at $v_{\rm{start}} = 5324$ \kmps and $v_{\rm{end}} = 5556$ \kmps (vertical cyan dotted lines). 
The minimum width at 50 per cent of the peak profile intensity is $W_{50} = 21 \pm 15$ \kmps (horizontal cyan solid double-arrow line) and the maximum width at 20 per cent of the peak profile intensity is $W_{20} = 150 \pm 22$ \kmps (horizontal cyan dashed double-arrow line).
The $W_{20}$ spans 23 channels whereas the $W_{50}$ spans only 4 channels revealing that the profile is steep and has a narrow and a broad component.
The average optical velocity resolution, within the profile edges, is $\Delta v = 6.8$ \kmps. 
The velocity centroid of the profile, $v_{\rm{HI}} = 5444 \pm 7$ \kmps, is defined as the mid velocity of $W_{20}$ (vertical blue dash-dot line and error-bar). 

The RMS noise, in all the line-free channels of the baseline subtracted line spectrum (\ie within 3940 \kmps $<v< 6961$ \kmps), is $\sigma_{S} = 3$ mJy.
We check the significance of the H{\sc i} detection in Fig. \ref{fig:HIspec}a, which displays the velocity range, 4345 \kmps $< v < $ 6556 \kmps.
The horizontal grey shaded region marks three times the spectral noise.
The flux density of the channel at the peak of the absorption profile, $S_{\rm{peak}} = -28 \pm 3$ mJy, is significant at a 8$\sigma$ level.
The flux density integrated between the profile edges, $S_{\rm{HI}} = -1.9 \pm 0.3$ Jy \kmps, is significant at a 6$\sigma$ level.
The profile area, \ie $S_{\rm{HI}}$, is shaded in blue within the spectrum.
The uncertainties on the peak and integrated flux densities, velocity centroid and widths of the H{\sc i} profile are calculated using equations from \citet{Koribalski2004}.

In Fig. \ref{fig:HIspec}a, we also study the kinematics of the H{\sc i} relative to that of ESO 137-G006 and the Norma cluster. $v_{\rm{HI}}$ is coincident with optical recessional velocity of the BCG, $v_{\rm{BCG}} = 5441 \pm 52$ \kmps \citep[vertical magenta dashed line and error-bar;][]{Woudt2008}.
However, both $v_{\rm{HI}}$ and $v_{\rm{BCG}}$ are redshifted compared to Norma's mean line-of-sight velocity, $v_{\rm{cluster}} = 4871 \pm 54$ \kmps \citep[vertical golden dotted line and error-bar;][]{Woudt2008}, and velocity dispersion, $\sigma_{v,\rm{Norma}} = 925$ \kmps \citep[vertical golden shaded region;][]{Woudt2008}.

The H{\sc i} absorption line reported in this paper is not found with the HIPASS data, consistent with the findings of \citet{Allison2014}. 
We verify that this is indeed the case, due largely to the effects of beam smearing (see appendix \ref{sec:HIPASS} for more details). 
The MeerKAT observations of ESO 137-G006, although more spatially resolved, too cannot be used to detect this absorption as they have poorer spectral resolution ($\sim 44$ \kmps) than our ATCA data \citep{Ramatsoku2020}. 

\subsection{H{\sc i} Properties}
\label{sec:NHI}

In this section, we investigate the density and temperature of the detected H{\sc i} absorption. 

\begin{figure*}
    \centering
    \includegraphics[width=\textwidth]{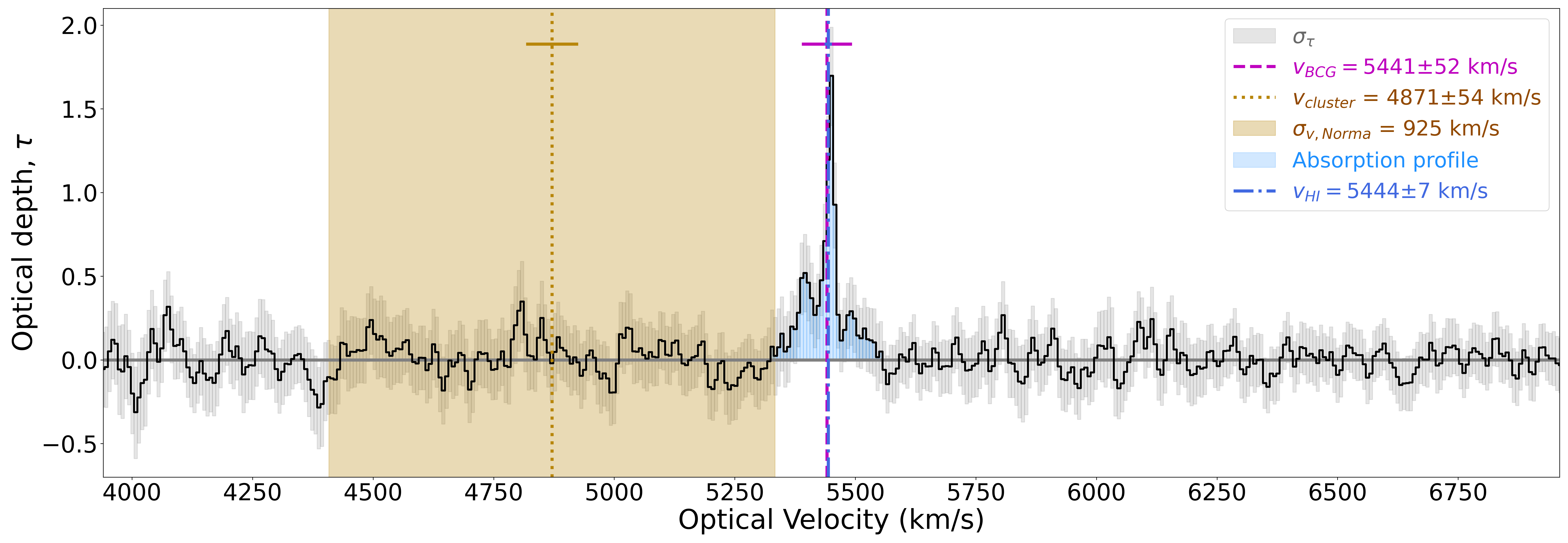}
    \caption{
    Optical depth and its uncertainty as a function of optical recessional velocity calculated as per equation \ref{eq:tau}, using the smoothed line and continuum spectra from the H{\sc i} detection region of the elliptical aperture towards ESO 137-G006's core. 
    }
    \label{fig:HItau}
\end{figure*}

\subsubsection{Optical depth of H{\sc i} absorption}

The optical depth, $\tau(v)$, as a function of velocity, $v$ in \kmps, is calculated as:
\begin{equation}
    \tau(v) = - \ln \left(1+ \frac{\Delta S(v)}{c_{f}S_{c}(v)} \right)
    \label{eq:tau}
\end{equation}

where $\Delta S (v)$ is the flux density from the absorption spectrum, $S_{c}(v)$ is the flux density from the continuum spectrum, and $c_{f}$ is the covering factor of the background source. 
We assume $c_{f} = 1$ throughout this paper.
The optical depth of the detected H{\sc i} and its uncertainty are presented in Fig. \ref{fig:HItau} as the black solid line and grey shaded region over a wide velocity range, 3940 \kmps $< v < $ 6961 \kmps. 
The peak absorption fraction, at $v_{\rm{peak}}$, is $(\Delta S/S_{c})_{\rm{peak}} = 0.82 \pm 0.40$, whereas the total absorption fraction, within the profile edges, is $(\Delta S/S_{c})_{\rm{total}} = 0.24\pm 0.03$.
Hence, the peak optical depth of the profile is $\tau_{\rm{peak}} = 1.7 \pm 0.3$ and the optical depth integrated between the profile edges is $\int\tau_{\rm{HI}} = 74 \pm 8$\kmps. 
The uncertainties of $\Delta S$ and $S_{c}$ have been added in quadrature to determine the uncertainty of $\tau$.

We use the velocity integrated optical depth to estimate the H{\sc i} absorption column density, in atoms cm$^{-2}$, using:
\begin{equation}
    \frac{N_{\rm{HI,abs}}}{\rm{cm^{-2}}} \approx 1.823 \times 10^{18} \times \frac{T_{\rm{spin}}}{\rm{K}} \int \tau_{\rm{HI}} \frac{dv}{\rm{km \, s^{-1}}}
    \label{eq:NHI}
\end{equation}
where the spin temperature (in Kelvin) of the gas, $T_{\rm{spin}}$, is assumed as the single excitation temperature in the partition function and the constant results from ignoring $n > 1$ states and assuming the Rayleigh-Jeans limit ($h\nu \ll kT$) and the known spontaneous emission coefficients for the H{\sc i} 21-cm transition \citep[see][section 2.1.1. for the complete derivation of equation \ref{eq:NHI}]{Allison2022}.
The integrated optical depth corresponds to an absorption column density of $N_{\rm{HI,abs}} \approx (1.3 \pm 0.2) \times 10^{20} \, T_{\rm{spin}} \, \rm{cm}^{-2}$.

\subsubsection{Constraining the spin temperature from emission non-detection}

The 21-cm transition in the hyper-fine splitting of the H{\sc i} n = 1 state has a very low transition probability \citep{Borthakur2010, Reeves2016}.
So, a substantially dense column ($N_{\rm{HI}} \geq 10^{19}$ cm$^{-2}$) is required to detect the H{\sc i} spectral line in absorption.
The column density is proportional to the spin temperature, $T_{\rm{spin}}$, which is the weighted harmonic mean of different thermal components of H{\sc i}. 
An estimate of $T_{\rm{spin}}$ can be used to constrain the densities and intensities of these different thermal components \citep[][and references therein]{Allison2022}. 
There are two stable phases of H{\sc i} in pressure equilibrium: the diffuse warm neutral matter (WNM) and dense cold neutral matter (CNM). 
In the CNM regions, spin temperature can be approximated as the kinetic temperature, 100 K $\leq T_{\rm{K}} \leq$ 1000 K, since collisions dominate the H{\sc i} hyper-fine level populations. 
On the other hand, in the WNM regions radiative excitation starts to dominate collisional excitation, increasing the spin temperature to 5000 K $\leq T_{\rm{K}} \leq$ 8000 K. 
This rise in temperature may be due to the presence of a strong continuum source, shocks, or other extreme conditions in the WNM \citep[\eg][]{Morganti2001, Morganti2005}. 
Hence to determine the physical phase of the H{\sc i}, we attempt to constrain the spin temperature. 

The H{\sc i} spin temperature can only be directly estimated from a detection in emission, not absorption \citep[][]{Allison2021}. 
However, since we did not detect the 21-cm line in emission, we can only provide an upper-limit on the spin temperature. 
This upper limit is estimated by deriving the H{\sc i} column density from an emission non-detection. 
To determine if the emission non-detection is meaningful, we calculate the upper-limit on H{\sc i} mass with it. 
The mass of H{\sc i} is related to its integrated emission flux, $S_{\rm{int}}$ in Jy \kmps, and distance, $d$ in Mpc, by:
\begin{equation} 
    \frac{M_{\rm{HI}}}{\mathrm{M}_{\odot}} = 2.36 \times 10^{5} \frac{S_{\rm{int}}}{\rm{Jy \, km \, s^{-1}}} \times \left(\frac{d}{\rm{Mpc}}\right)^{2}
    \label{eq:MHI}
\end{equation}
The detection limit, in the line spectrum of the H{\sc i} detection region, is 3$\sigma_{S} \approx 10$ mJy.
To estimate $S_{\rm{int}}$, we integrate this limit as a top-hat function over an approximation of the emission line width ($\tilde{W}$).
$\int \frac{dv}{\rm{km.s^{-1}}} = \tilde{W} \approx 150$\kmps is assumed in previous studies of the Norma cluster \citep{Vollmer2001a, Schroder2009}. 
This results in the upper limits $S_{\rm{int}} \leq 1.5$ Jy \kmps and $M_{\rm{HI}} \leq 1.7 \times 10^{9} \, \mathrm{M}_{\odot}$. 
This upper limit is in line with the expected H{\sc i} mass content of ETGs in the local Universe.
For example, \citet{Serra2012} has shown that only 5\% of the ETGs observed by the $\rm{ATLAS^{3D}}$ project have global H{\sc i} masses greater than $1.7 \times 10^{9} \, \mathrm{M}_{\odot}$.

Since the emission non-detection is meaningful, we estimate the H{\sc i} emission column density by integrating the brightness temperature ($T_{\rm{B}}$) over $\tilde{W}$ as:
\begin{equation}
    \frac{N_{\rm{HI,em}}}{\rm{cm^{-2}}} \approx 1.823 \times 10^{18} \int \frac{T_{\rm{B}}}{\rm{K}} \frac{dv}{\rm{km \, s^{-1}}}
    \label{eq:NHITb}
\end{equation}
The above equation is derived assuming the background radiation is negligible and approximating $T_{\rm{S}}\int \tau_{\rm{HI}} dv \approx \int T_{\rm{B}} dv$ \citep[for details, see][section 15]{Meyer2017}.
The velocity integrated $T_{\rm{B}}$ is the integrated temperature an extended black body needs to have to produce the observed integrated emission flux, $S_{\rm{int}}$, in the Rayleigh-Jeans limit \citep[see][ section 14]{Meyer2017}.
Assuming a Gaussian beam of solid angle $\Omega_{\rm{bm}}= (\pi \theta_{a}\theta_{b})/(4ln(2))$, we obtain the following equation for the velocity integrated $T_{\rm{B}}$:
\begin{equation}
    \int \frac{T_{\rm{B}}}{\rm{K}} \frac{dv}{\rm{km \, s^{-1}}} \approx 1.360 \times \left(\frac{\lambda}{\rm{cm}}\right)^{2} \times \frac{1}{\theta_{a}\theta_{b}} \times \frac{S_{\rm{int}}}{\rm{mJy \, km \, s^{-1}}}
    \label{eq:TbSint}
\end{equation}
where $\theta_{a}=9$ arcsec and $\theta_{b}= 7$ arcsec are the FWHM of the synthesised beam and $\lambda = 21.1$cm for the H{\sc i} line.
Using the constrained $S_{\rm{int}}$ in equation \ref{eq:TbSint}, we obtain the upper limits $T_{\rm{B}} \leq 1.4 \times 10^{4}$ K and $N_{\rm{HI,em}} \leq 2.6 \times 10^{22}$ cm$^{-2}$. 
We then divide this emission column density estimate by the absorption column density estimate, \ie $N_{\rm{HI,abs}} \approx (1.3 \pm 0.2) \times 10^{20} \, T_{\rm{spin}}$ cm$^{-2}$, to get the spin temperature upper limit, $T_{\rm{spin}} \leq 194$ K.

$T_{\rm{B}}$, thus $T_{\rm{spin}}$, can also be constrained by working backwards from the stellar mass of ESO 137-G006 estimated in \citet{Vaddi2016}.
Scaling-relations of H{\sc i} gas fraction with stellar mass, of other galaxies in the local universe \citep[\eg][]{Catinella2018}, can be used to estimate $M_{\rm{HI}}$, thus $S_{\rm{int}}$ (instead of 3$\sigma_{S}$). 
However, the upper limit on $T_{\rm{spin}}$ from the non-detection of emission in ESO 137-G006 provides a tighter constraint. 
Therefore, we confirm that $T_{\rm{spin}} \leq 194$ K is consistent with the detected H{\sc i} being in the cold neutral phase.
In the next section, we discuss the implications of detecting dense cold H{\sc i} gas against ESO 137-G006's core that is likely a young compact source.

\section{Discussion}
\label{sec:Discussion}
In this section, we explore how the detected H{\sc i} fits in with the evolution of ESO 137-G006 and discuss the limitations of our study.

\subsection{Association of the H{\sc i} absorption with ESO 137-G006}
\label{sec:CR}

From the spatial and spectral analysis, we find it likely that the H{\sc i} detected in absorption against ESO 137-G006's core is associated with this BCG. 
The H{\sc i} detection region is not only spatially located within the BCG's 2MASS isophotal diameter but the H{\sc i} profile's velocity centroid ($5444 \pm 7$\kmps) is also spectrally coincident with the BCG's recessional velocity \citep[$5441 \pm 52$ \kmps, ][]{Woudt2008}. 
Moreover, given the width and the shape of its absorption profile, the H{\sc i} appears to be a part of settled gas within ESO 137-G006.
\citet{Gereb2015}, hereafter \citetalias{Gereb2015}, studied the H{\sc i} absorption profiles of 32 radio-loud galaxies similar to ESO 137-G006 (\ie $S_{1.4 \, \rm{GHz}} > 50$ mJy). 
\citetalias{Gereb2015} found, like our detection against ESO 137-G006, that 10 of the H{\sc i} profiles were narrow ($ W_{50} < 100$ \kmps and $W_{20} < 200$ \kmps) and symmetric ($< 50$ \kmps difference between the profile's velocity centroid and peak). 
Of these narrow and symmetric profiles, the ones that were coincident with the systemic velocity of their background radio galaxies were indicative of the H{\sc i} being settled in large-scale disks. 
This is consistent with the detected H{\sc i} absorption profiles in the $\rm{ATLAS^{3D}}$ sample of ETGs with typical $ W_{50} < 80$ \kmps \citep[][\citetalias{Gereb2015}]{Serra2012}.

In brief, given its location, velocity, width and shape, the detected H{\sc i} absorption feature in this study is unlikely to be an outflow in our line of sight and more likely to be settled gas within ESO 137-G006.
However, the kinematics of the H{\sc i} profile are complex relative to the velocity dispersion of the Norma cluster. 
So, the origin of the H{\sc i} within the BCG could either be left-over gas reservoirs from the formation stages or newly acquired gas reservoirs from interactions in evolutionary stages. 

\subsection{Cold Gas Properties of Early-Type Galaxies}

From previous studies, we know that ESO 137-G006 is an ETG  \citep[\eg][]{Vaddi2016}. 
Using sensitive and targeted observations, \citet{Serra2012} and \citet{Davis2019} detected, in emission, a significant fraction of ETGs with dense cold gas reservoirs albeit in smaller amounts than those of similarly massive late-type counterparts \citep{Saintonge2022}. 
We detect small amounts ($M_{\rm{HI}} \leq 1.7 \times 10^{9} \, \mathrm{M}_{\odot}$) of dense ($N_{\rm{HI}} \approx (1.3 \pm 0.2) \times 10^{20} \, T_{\rm{spin}}$ cm$^{-2}$) and cold ($T_{\rm{spin}} \leq 194$ K) H{\sc i} gas, intrinsic to ESO 137-G006. 
This detection is consistent with the cold gas properties of other ETGs, detected in absorption \citep[\eg][]{Morganti2006, Oosterloo2010} and emission \citep[\eg][]{Serra2012, Davis2019}. 

Given this detection of intrinsic dense cold atomic gas, it may be possible that ESO 137-G006 also possesses detectable cold molecular gas. 
\citet{Davis2019} found molecular carbon-monoxide (CO) gas reservoirs in 22.4\% of ETGs in the local universe from the MASSIVE and $\rm{ATLAS^{3D}}$ surveys.
This was the case regardless of the galaxies' mass, size, position on the fundamental plane, and local environment. 
Moreover, \citet{Serra2012} observed that ETGs with small H{\sc i} discs were forming molecular gas at efficiencies comparable to spiral galaxies, despite having lower H{\sc i} column densities. 
\citet{Serra2012} also found on-going star-formation activity in 70\% of ETGs with central H{\sc i}. 
However, the measurement or upper limit of molecular content towards ESO 137-G006 has not been published to date. 
\citet{Vaddi2016} estimate a low star-formation rate (SFR = $1.54\times10^{-3}\,\mathrm{M}_{\odot}\,\rm{yr^{-1}}$) in ESO 137-G006 from far-ultraviolet luminosities corrected for contribution from the evolved stellar populations. 
This low SFR in ESO 137-G006 is consistent with the BCG's Wide-field Infrared Survey Explorer (WISE) mid-infrared colour-colour properties \citep[\eg][]{Wright2010, Jarrett2017}, which suggest that the BCG is passive. 
This means that the majority of the mid-infrared emission from the BCG is due to radiation from its older stellar population.

In summary, the mass, density and temperature of the H{\sc i} intrinsic to ESO 137-G006 is as expected from what is observed in other ETGs. 
The high H{\sc i} column density indicates that the BCG might also possess some molecular gas, but not in amounts enough to feed significant star-formation.

\subsection{Intrinsic Cold Gas in Brightest Cluster Galaxies}

We know that ESO 137-G006 is the Norma cluster's central BCG \citep{Boehringer1996, Sun2007, Woudt2008, Sun2009a}. 
So, the detected H{\sc i} gives us more information regarding the gas content in the centre of this very massive cluster.
Nevertheless, it is difficult to check if the H{\sc i} properties of ESO 137-G006 are consistent with the gas properties of other BCGs as a comprehensive search for H{\sc i} in absorption towards BCGs has not been conducted to date.
A sufficiently bright background radio source, such as a radio-loud AGN \citep[$L_{1.4 \, \rm{GHz}} \geq 10^{24} \, \rm{W \, Hz^{-1}}$, \eg][]{Sun2009a}, is required to detect H{\sc i} in absorption. 
But not all BCGs are radio-loud galaxies like ESO 137-G006.
This makes detecting the H{\sc i} absorption line difficult, even against BCGs that have the most massive or the most dense gas reservoirs.

As introduced earlier, the Coma cluster is thought to be a more distant ($z$ = 0.023) and virialised counterpart of the Norma cluster \citep{Kraan-Korteweg1996}.
Like Norma, Coma also hosts two dominant cD type galaxies in its core.
These galaxies, NGC 4889 and NGC 4874, have fainter 1.4 GHz luminosities ($L_{1.4 \, \rm{GHz}} = 10^{21.08} \, \rm{and} \, 10^{23.35} \, \rm{W \, Hz^{-1}}$, respectively) than that of ESO 137-G006 \citep[$L_{1.4 \, \rm{GHz}} = 10^{25.38} \, \rm{W \, Hz^{-1}}$,][]{Sun2007}. 
As a consequence, Coma's cD galaxies do not have bright enough continuum sources to enable the detection of H{\sc i} in absorption.
However, the cD galaxies have also not been detected with H{\sc i} in emission \citep[$M_{\rm{HI}} < 10^{7.43} \, \mathrm{M}_{\odot}$,][]{Healy2021a}. 
\citet{Vollmer2001a} pose that Coma's cD galaxies may in fact be H{\sc i}-deficient either as a result of evaporation, caused by conduction from the surrounding hot intergalactic medium (IGM), or due to ram-pressure stripping as they moved through the dense intra-cluster medium (ICM) in the early stages of cluster evolution.

Despite the difficulties of making the detection in absorption, multiple BCGs have been detected with H{\sc i} attributed to the cool-core of their host clusters or groups \citep[][and references therein]{Hogan2014, Morganti2018}.
Some examples of these, from closest to farthest, include:
PKS 0316+413 (NGC 1275) in the Perseus Cluster \citep[ACO 426, $z$ = 0.018;][]{DeYoung1973, Jaffe1988, Sijbring1989, Jaffe1990, Momjian2002, Chandola2013, Moss2017, Grasha2019}, 
BCG of 2A0335+096 \citep[$z$ = 0.035;][]{McNamara1990}, 
PKS 1519+07 (NGC 5920) in MKW3s \citep[$z$ = 0.044;][]{McNamara1990}, 
UGC 2489 in A407 \citep[$z$ = 0.045;][]{Shulevski2015}, 
PKS 0915-11 (Hydra-A) in A780 \citep[$z$ = 0.055;][]{Taylor1996}, 
Cygnus-A \citep[$z$ = 0.056;][]{Conway1995, Struve2010}, 
PKS 1346+268 in A1795 \citep[$z$ = 0.064;][]{vanBemmel2012}, 
PKS 2322-123 in A2597 \citep[$z$ = 0.082;][]{ODea1994, Taylor1999}, 
PKS 1555-140 \citep[$z$ = 0.097;][]{Curran2006}, 
PKS 1353-341 \citep[$z$ = 0.223;][]{Veron-Cetty2000}. 

The H{\sc i} profiles are detected against the strong central AGN activity of these cool-core cluster/group BCGs.
The profiles have widths between 100-1000 \kmps and column densities between $10^{21-23} \rm{cm}^{-2}$, similar to our detection against Norma's BCG. 
However, unlike our detection, most of these detections have a broad associated component and a narrow redshifted/blueshifted component \citep{Morganti2018}. 
These detections are seen as evidence for a surrounding `cooling flow' being responsible for the presence of H{\sc i}, which feeds the AGN activity of these BCGs \citep{Hogan2014}.
But it is difficult to draw direct comparisons between the environment of these BCGs and that of ESO 137-G006.
For example, NGC 1275 in Perseus is the closest (albeit further than ESO 137-G006) cool-core cluster BCG detected with H{\sc i} and it has a cluster environment that is as massive and complex as that of ESO 137-G006 \citep{Boehringer1996, Woudt1998, Sun2009}.
However, firstly, NGC 1275 is located in Perseus' large cool-core, whereas Norma has a rarer and smaller cool-core, of the luminous corona type, associated with ESO 137-G006's strong AGN \citep{Sun2007, Sun2009, Nishino2012}.
Secondly, the 1.4 GHz luminosity of NGC 1275's AGN is only 65\% that of ESO 137-G006's AGN \citep{Sun2009}.
Lastly, despite the presence of strong AGN feedback, NGC 1275 is exceptionally found with cold molecular gas and a high SFR ($\sim 41 \, \rm{Myr}^{-1}$) near its core \citep{Fraser-McKelvie2014}.
Thus, NGC 1275 cannot be used to argue for a cooling flow origin for the cold gas that has been discovered in ESO 137-G006.

In short, our findings are only partly consistent with the H{\sc i} absorption properties detected towards other cool-core cluster BCGs and we cannot confirm that the H{\sc i} in ESO 137-G006 is present due to Norma's central cooling flow.

\subsection{Sub-cluster merger in Norma}

The X-ray distribution of Norma is not spherically symmetric, which indicates that Norma is not in dynamic equilibrium and is possibly ongoing a sub-cluster merger \citep{Boehringer1996, Tamura1998}. 
The observed X-ray sub-cluster, due to its compactness, suggests that the merger has commenced recently and not progressed very far \citep{Boehringer1996}.
Norma's elliptical/spheroidal population is relaxed and the spiral/irregular population is distributed in various substructures.
But both of these galaxy populations are aligned along the nearby dominant large-scale structure, the Norma Wall \citep{Woudt2008}.
So, the sub-cluster merger has been associated with the formation of this large-scale structure \citep{Nishino2012}.

The X-ray sub-cluster is aligned, in projection, with the 843-MHz radio continuum emission of ESO 137-G006's WAT \citep{Jones1996, Woudt2008}, placing this BCG at the leading edge of this sub-cluster merger. 
Furthermore, there is an observed line-of-sight velocity offset between the mean cluster velocity ($4871 \pm 54$ \kmps) and that of ESO 137-G006 \citep[$5441 \pm 52$ \kmps;][]{Woudt2008}. 
So, it is possible that ESO 137-G006's WAT is bent due to the ram-pressure exerted on it as the BCG moves, as part of the sub-cluster, through the dense ICM \citep{Owen1976, Sakelliou2000}.
Altogether, these observations suggest that ESO 137-G006 may have recently merged with the cluster core as a consequence of the merger interactions, thereby retaining some H{\sc i} from its formation stages.

In summary, it is possible that some cool-core BCGs, like ESO 137-G006, are young radio galaxies that retain small amounts of cold gas in their core despite internal evaporation or stripping from their sub-cluster environments.
Compared to other central dominant galaxies of low redshift ($z$ < 0.03) massive ($M \sim 10^{15}\,\mathrm{M}_{\odot}$) clusters such as Coma and Perseus, Norma's BCG is the only one with detected intrinsic H{\sc i} gas that is not associated with a cluster cooling flow.
This suggests that detectable H{\sc i} gas associated with the radio core of massive (\ie $M \sim 10^{15}\,\mathrm{M}_{\odot}$) cluster BCGs is likely to be less common.
Our results are consistent with the scenario that ESO 137-G006 may have recently moved into the cluster centre as opposed to being Norma's original BCG at the bottom of its potential well.

\subsection{Limitations and Improvements}
\label{sec:Lim}

Imaging these observations with the standard imaging process is challenging as the targeted region has multiple sources of strong extended continuum that are within and adjacent to each ATCA pointing, which has a wider-field nature at 1.4 GHz relative to higher frequencies.
In addition, our observations are significantly affected by sidelobes due to the insufficient $uv$-coverage and limited surface brightness sensitivity of ATCA's 6A array configuration, used for achieving the highest spatial resolution.
The limited $uv$-coverage and significant sidelobe effects result in a baseline ripple in the line spectra after imaging.
Moreover, the detection of Norma's galaxy members has been incomplete to date due to its location near the Galactic plane - the optical and infrared zone of avoidance (ZOA). 
Thus, the estimates of Norma's mean cluster velocity and velocity dispersion remain highly uncertain \citep{Woudt2008, Schnitzeler2010}.
Constraining the kinematics of this cluster could in turn constrain the movement of ESO 137-G006 with respect to it and thus improve our understanding of the role of H{\sc i} in the BCG's evolution. 

Future H{\sc i} surveys that can see beyond the ZOA will reveal much of the Universe that remains largely understudied at shorter wavelengths. 
With the use of the Australian Square Kilometre Array Pathfinder (ASKAP), the upcoming Wide-field ASKAP L-band Legacy All-sky Blind surveY \citep[WALLABY,][]{Koribalski2020a} can help detect H{\sc i} emission in and around the Norma cluster. 
These H{\sc i} emission detections could then be used to better constrain Norma's kinematics and merger history. 
Additionally, other future Square Kilometre Array (SKA) pathfinders surveys \citep[\eg][]{Gupta2021, Allison2022} are deeper, have more complete $uv$-coverage, and have greater angular resolution than our ATCA data.
So, more detailed solutions for continuum subtraction of multiple bright extended sources (\eg peeling) can be applied to the SKA pathfinders' data in the future.
These surveys will also be equipped to conduct a comprehensive survey of cold gas in the BCGs of massive clusters. 
Therefore, future studies with such surveys may be able to quantify the fraction of massive cluster BCGs that are accreted into the cluster centre at a later stage of cluster evolution.

\section{Conclusions}
\label{sec:End}
Using ATCA observations of high resolution and sensitivity, we have detected H{\sc i} associated with ESO 137-G006, a BCG of the Norma cluster. 
The H{\sc i} absorption source is located at the coordinates of the BCG (R.A. (J2000) = 16:15:03.7 and Dec (J2000) = $-60$:54:24.9), and is constrained within the BCG's 2MASS isophotal diameter.
The velocity centroid ($5444 \pm 7$ \kmps) of the H{\sc i} absorption profile is coincident with the BCG's velocity \citep[$5441 \pm 52$ \kmps][]{Woudt2008}. 
Moreover, the detected profile is symmetric and narrow ($W_{20} = 150\pm22$\kmps), suggesting the gas may be settled in a disk within ESO 137-G006 \citep{Gereb2015}.
The spectral index ($\alpha = 0.3 \pm 0.2$) of the likely compact (< 3 kpc) radio core ($65 \pm 18$ mJy) in the background of this detection suggests that the AGN activity in ESO 137-G006 may have restarted. 
However, we do not have the required observations to investigate if the H{\sc i} is feeding the AGN. 

The estimated density of the H{\sc i} absorption column is $N_{\rm{HI}} \approx (1.3 \pm 0.2) \times 10^{20}\,T_{\rm{spin}}$ cm$^{-2}$. 
We constrain the spin temperature, $T_{\rm{spin}} \leq 194$ K, from the upper limit of emission non-detection ($M_{\rm{HI}} \leq 1.7 \times 10^{9}\,\mathrm{M}_{\odot}$). 
The low spin temperature suggests that the absorption column is in the cold neutral phase. 
Whereas the high column density, consistent with the cold gas properties of other ETGs, suggests that small amounts of cold molecular gas could also be present in this BCG's core without triggering significant star-formation \citep{Morganti2006, Oosterloo2010, Serra2012, Davis2019}. 

A `cooling flow' origin of the H{\sc i} within ESO 137-G006 cannot be confirmed since the detection is only partly consistent with the properties of other cool-core cluster BCGs \citep{Hogan2014, Morganti2018}. 
On the other hand, it is possible that ESO 137-G006 has recently arrived into Norma's core.
So, the BCG may have retained the H{\sc i} from its formation stages, or attained it from interactions, in the process of moving into the cluster core as the leading-edge of an ongoing sub-cluster merger. 
The alignment of ESO 137-G006's WAT with Norma's X-ray sub-cluster and the BCG's significant line-of-sight velocity offset from that of the cluster mean \citep{Boehringer1996, Jones1996, Woudt2008} support this merger scenario. 

This study is limited by our simple imaging process and the Norma cluster's position near the Galactic plane. 
With data from future all-sky surveys of the SKA pathfinders \citep[\eg][]{Koribalski2020a, Allison2022, Gupta2021}, a more complete and detailed study of Norma can be conducted.
A comprehensive census of H{\sc i} associated with other BCGs can also be taken.
Such a census would provide the high statistics needed to determine if the presence of cold-gas can help identify the BCGs that have been accreted into the cluster at a later stage in cluster evolution.


\section*{Acknowledgements}
We thank the referee for constructive comments that improved this manuscript.
OIW and BSK were a part of the ATCA observation proposal and observed the data. 
MS lead the data reduction and analysis with advice from OIW and Lister Staveley-Smith. 
MS, LC and OIW are responsible for the interpretation of the results into the discussion. 

LC is the recipient of an Australian Research Council Future Fellowship (FT180100066) funded by the Australian Government. 
Parts of this research were conducted by the Australian Research Council Centre of Excellence for All Sky Astrophysics in 3 Dimensions (ASTRO 3D), through project number CE170100013.

The Australia Telescope Compact Array is part of the Australia Telescope National Facility (\url{https://ror.org/05qajvd42}) which is funded by the Australian Government for operation as a National Facility managed by CSIRO. 
We acknowledge the Gomeroi people as the traditional owners of the Observatory site. 
This paper includes archived data obtained through the Australia Telescope Online Archive (\url{http://atoa.atnf.csiro.au}).

\section*{Data Availability}
The reduced observations are available on the CSIRO Data Access Portal, at \url{https://doi.org/10.25919/1842-k647}.


\bibliographystyle{mnras}
\bibliography{refs_better.bib} 


\appendix
\section{Observation Table}
\label{sec:ObsTable}

Table \ref{tab:Obs} presents the log of Project C2891 for observations made with the 6A array configuration and the 64-MHz zoom-mode CABB configuration. 
For the two calibrators and seven source pointings used in this study, table \ref{tab:Obs} logs the number of observation cycles, with their respective time-lengths, and the total observation time. 
Due to RFI, we exclude the first observation cycle of the primary calibrator and extend some observation cycles of pointings 12, 17 and 18.

\begin{table*}
\centering
\caption{Log of the ATCA observations used.}
\label{tab:Obs}
\begin{tabular}{rlll}
\textbf{Source}      & \textbf{Name} & \textbf{Observation Cycles}                     & \textbf{Total Time Observed} \\ \hline \hline
Primary calibrator   & 1934-638      & 1 (16m50s; at the start) + 1 (7m4s; at the end) & 24m30s                       \\ \hline
Secondary calibrator & 1613-586      & 1 (7m) + 14 (4m20s) + 1 (1m10s) + (5m20s)       & 1h01m10s                     \\ \hline
Pointing             & 12            & 11 (2m) + 1 (1m18s)                             & 23m18s                       \\ \hline
Pointing             & 13            & 11 (2m)                                         & 22m00s                       \\ \hline
Pointing             & 14            & 11 (2m)                                         & 22m00s                       \\ \hline
Pointing             & 17            & 11 (2m) + 1 (2m10s)                             & 24m10s                       \\ \hline
Pointing             & 18            & 8 (2m) + 1 (2m8s) + 2 (2m)                      & 22m08s                       \\ \hline
Pointing             & 19            & 10 (2m)                                         & 20m00s                       \\ \hline
Pointing             & 23            & 10 (2m)                                         & 20m00s                       \\ \hline
\end{tabular}
\end{table*}

\section{Continuum Subtraction}
\label{sec:ContSub}

Table \ref{tab:UVLINorder} presents the order of the polynomial fit to the continuum emission in each of the pointing's $uv$-data. 
These are the lowest orders of the polynomial fit required to represent the individual continuum spectrum from each pointing's $uv$-data.

\begin{table}
\centering
\caption{The order of polynomial fit to the continuum emission to subtract it from the flagged and calibrated $uv$-data of each pointing.}
\label{tab:UVLINorder}
\begin{tabular}{cc}
\textbf{Pointing} & \textbf{Order of polynomial subtracted} \\\hline \hline
12 & 3 \\\hline
13 & 2 \\\hline
14 & 3 \\\hline
17 & 4 \\\hline
18 & 1 \\\hline
19 & 3 \\\hline
23 & 3 \\\hline
\end{tabular}
\end{table}

\section{Smoothing and Baseline Ripple Removal}
\label{sec:OGspectra}

\begin{figure*}
    \centering
    \includegraphics[width=\textwidth]{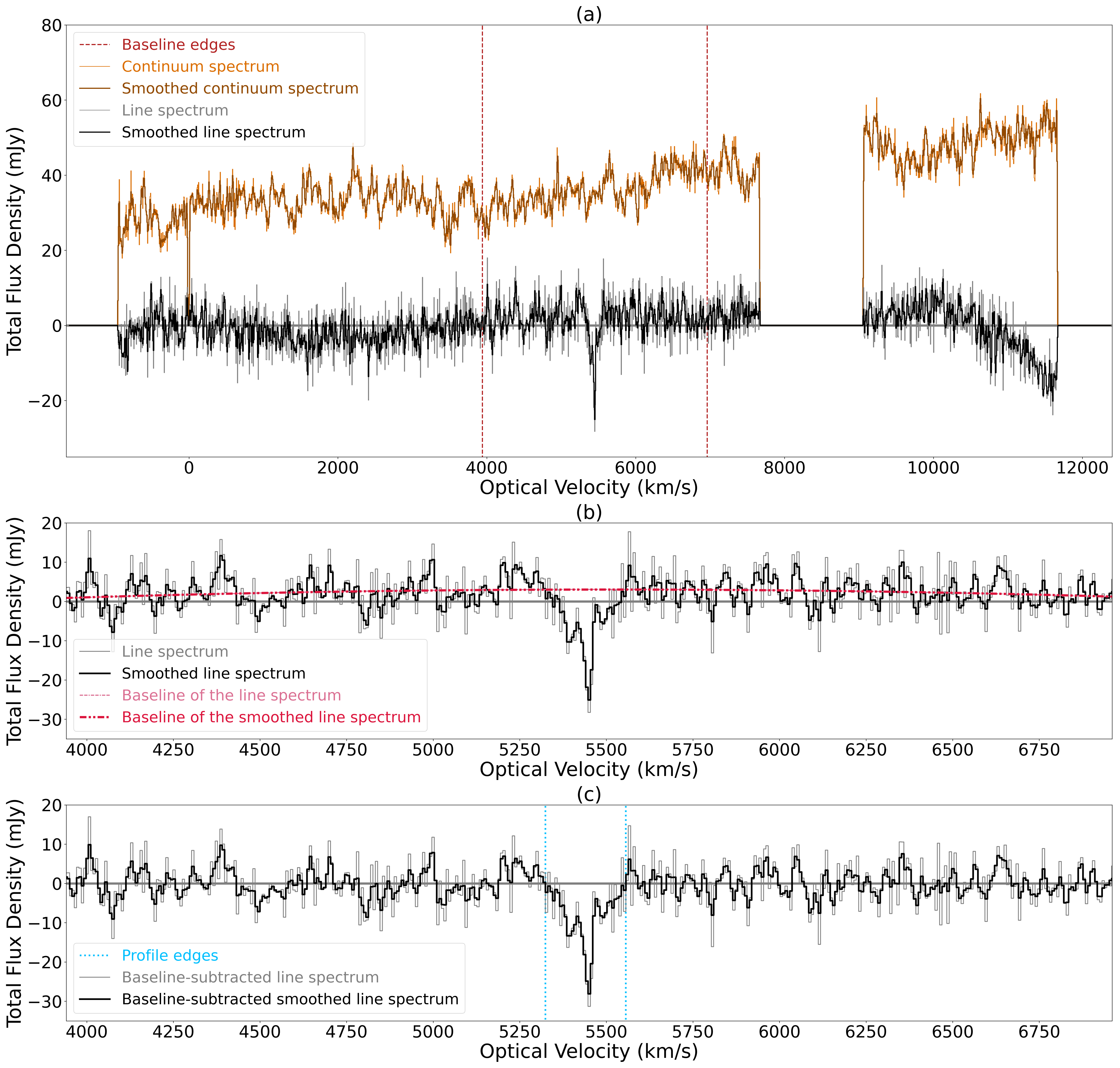}
    \caption{The spectra from the elliptical aperture are smoothed by a Hanning function of kernel = 3 and subtracted by a second order polynomial fit between the vertical red dashed lines.
    The line spectrum without (thin grey) and with (thick black) smoothing and the continuum spectrum without (thin orange) and with (thick brown) smoothing are displayed.
    The baseline for the line spectrum without smoothing (thin pink dash-dash-dot line) is the same as the baseline for the line spectrum with smoothing (thick red dash-dot-dot line).
    Between both the baseline subtracted line spectra, the smoothed profile has clearer edges (vertical cyan dotted lines). 
    But there are no significant differences between the H{\sc i} properties of the spectra with and without smoothing.
    }
    \label{fig:OGspec}
\end{figure*}

Smoothing the spectra by a Hanning function of kernel = 3, helps identify the profile edges more clearly but does not significantly affect the H{\sc i} properties. 
In Fig. \ref{fig:OGspec}, we contrast the spectra, extracted form the H{\sc i} detection region, without and with smoothing for completeness and reproducibility. 
The smoothed line spectrum (thick black line) is overlaid on the line spectrum without smoothing (thin grey line).
In Fig. \ref{fig:OGspec}a, the smoothed continuum spectrum (thick brown line) is overlaid on the continuum spectrum without smoothing (thin orange line).

The residual baseline ripple, seen in Fig. \ref{fig:OGspec}a over the full bandwidth of the observations (\ie $-$1610 \kmps $< v < $ 12381 \kmps), is removed by subtracting the baseline in the velocity range around the peak of the H{\sc i} absorption profile. 
We fit a second order polynomial baseline to the line spectrum over all line-free channels in the velocity range, 3940 \kmps $<v< 6961$ \kmps (see Fig. \ref{fig:OGspec}b).
The resulting second order polynomial is given by equation \ref{eq:ELPS_baseline}, which adequately models any curvature in the baseline within this velocity range without over fitting the spectrum.
A polynomial is fit to both line spectra (without and with smoothing) and the resulting baselines are the same (\ie equation \ref{eq:ELPS_baseline}) regardless of smoothing.
The baseline subtracted line spectra are shown in Fig. \ref{fig:OGspec}c.

\begin{equation}
    \frac{\Delta S}{\rm{mJy}} = -9\times10^{-7}\left(\frac{v}{\rm{km \, s^{-1}}}\right)^2+1\times10^{-2}\left(\frac{v}{\rm{km \, s^{-1}}}\right)-24 
    \label{eq:ELPS_baseline}
\end{equation}

The profile edges are measured at the channels where the profile meets or exceeds the 0 mJy line in the smoothed and baseline subtracted line spectrum.
To calculate the H{\sc i} properties, the same profile edges are assumed for both the line spectra without and with smoothing.
All the H{\sc i} properties calculated in this paper are summarised in table \ref{tab:HIprop}.
Column 2 and 3 of this table provide a direct comparison between the H{\sc i} properties of the two sets of spectra presented in Fig. \ref{fig:OGspec}.
Both sets of spectra recover statistically consistent H{\sc i} properties.
Only the smoothed and baseline subtracted line spectrum is presented in section \ref{sec:Results}.

\section{H{\sc i} Detection Region}
\label{sec:BoxSpectra}

\begin{figure*}
    \centering
    \includegraphics[width=\textwidth]{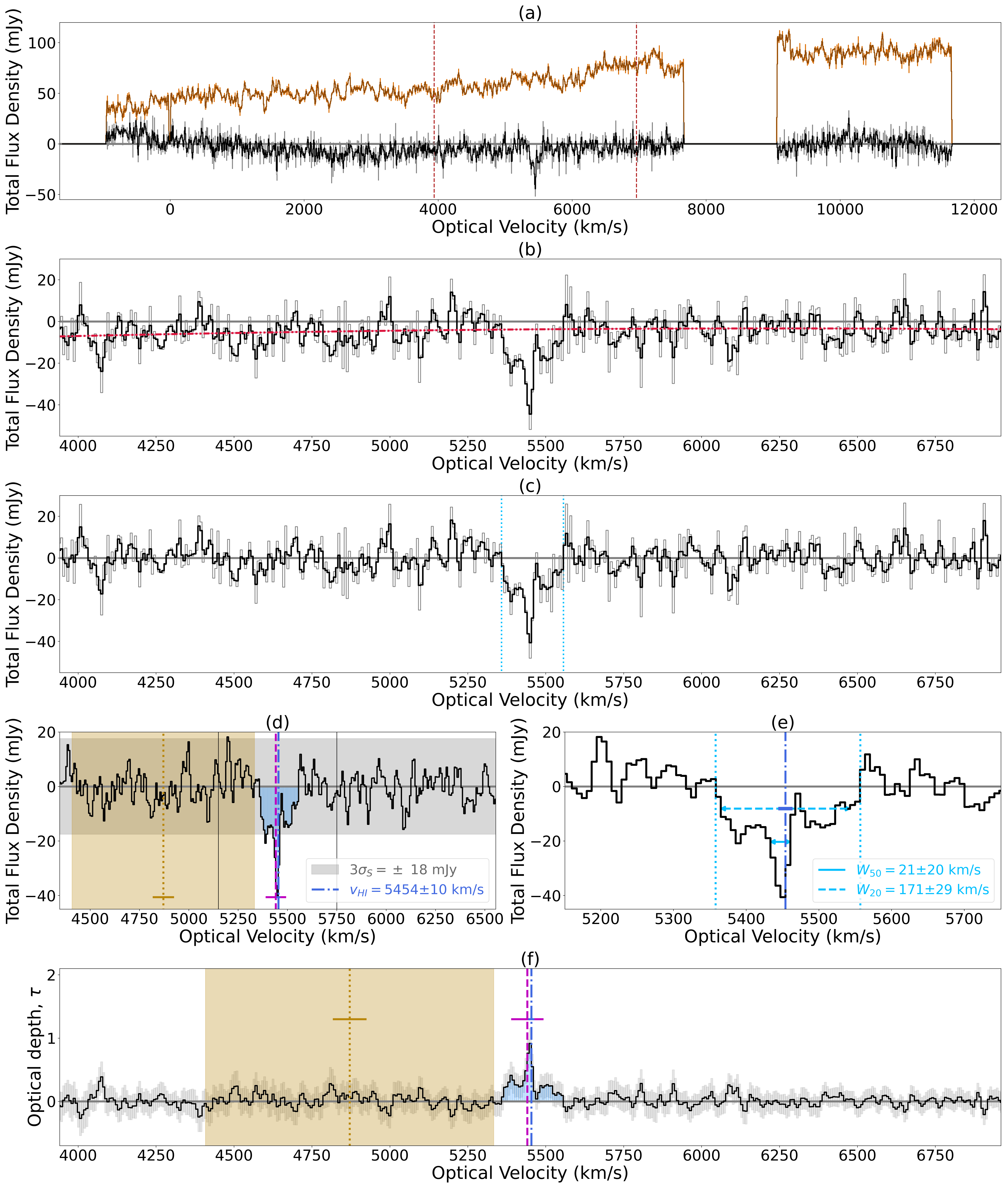}
    \caption{H{\sc i} spectra extracted from the ATCA data over a square aperture of length 22 arcsec centred at R.A. (J2000) = 16:15:04 and Dec (J2000) = $-60$:54:26.
    The effect of Hanning smoothing on the line and continuum spectra, over the full bandwidth, is shown in panel a (like Fig. \ref{fig:OGspec}a). 
    Panels b and c (like Fig. \ref{fig:OGspec}b and c) show that the second order polynomial baseline fit to the line spectra and the baseline subtracted spectra respectively.
    The H{\sc i} profile in the smoothed line spectrum is fit in panel e and is compared to the velocities of the host BCG and cluster in panel d (like Fig. \ref{fig:HIspec}). 
    The optical depth of the smoothed H{\sc i} profile is presented in panel f (like Fig. \ref{fig:HItau}).
    Overall, this figure shows that the H{\sc i} detection region does not significantly affect the H{\sc i} properties.
    }
    \label{fig:BoxPlot}
\end{figure*}

The H{\sc i} properties do not vary significantly if the spectra are extracted from the square aperture of length 22 arcsec rather than the H{\sc i} detection region reported in this study \ie an elliptical aperture of major and minor axis 18 arcsec and 14 arcsec respectively.
To showcase this, we extract the line and continuum spectra, from their respective spectral and continuum cubes, over the square aperture centred at R.A. (J2000) = 16:15:04 and Dec (J2000) = $-60$:54:26.
This square aperture covers ESO 137-G006's core and, whilst bigger than the H{\sc i} detection region, is small enough to exclude any continuum structure from ESO 1370-G006's WAT lobes.
We repeat the same analysis detailed in section \ref{sec:Results} on the spectra from the square aperture and present our results in Fig. \ref{fig:BoxPlot} and table \ref{tab:HIprop}.
All the spectra, baselines, shaded regions and vertical and horizontal lines in Fig. \ref{fig:BoxPlot} have the same colour and style as their respective comparative pair figure (\ie Fig. \ref{fig:OGspec}, \ref{fig:HIspec} or \ref{fig:HItau}) displaying spectra from the elliptical aperture.
Only the lines and shaded regions that have changed in value (with respect to Fig. \ref{fig:OGspec}, \ref{fig:HIspec} and \ref{fig:HItau}) are identified in the legend of Fig. \ref{fig:BoxPlot}.

Fig. \ref{fig:BoxPlot}a, b and c, replicate the analysis in Fig. \ref{fig:OGspec} to show the effect of smoothing the line and continuum spectra by a Hanning function with kernel$=3$.
To confirm smoothing does not affect the H{\sc i} properties, we perform the analysis on both the sets of spectra: with smoothing and without smoothing.
As in section \ref{sec:OGspectra}, we also remove the residual baseline ripple seen in Fig. \ref{fig:BoxPlot}a, by subtracting a second order polynomial baseline fit within the velocity range, 3940 \kmps $<v< 6961$ \kmps (see Fig. \ref{fig:BoxPlot}b). 
The resulting second order polynomial is given by equation \ref{eq:SQ_baseline} and is the same regardless of smoothing. 
The change in the baseline about 0 mJy depending on the aperture (\ie between Fig. \ref{fig:OGspec}b and Fig. \ref{fig:BoxPlot}b) is a result of the non-Gaussian noise properties caused due to the limited $uv$-coverage and significant sidelobe effects (see section \ref{sec:Lim}).
The baseline subtracted line spectra are shown in Fig. \ref{fig:BoxPlot}c.

\begin{equation}
    \frac{\Delta S}{\rm{mJy}} = -7\times10^{-7}\left(\frac{v}{\rm{km \, s^{-1}}}\right)^2+9\times10^{-3}\left(\frac{v}{\rm{km \, s^{-1}}}\right)-32 
    \label{eq:SQ_baseline}
\end{equation}

The profile is identified within the velocity range, 5358 \kmps $< v <$ 5557\kmps, which covers 5 less channels than the profile from the elliptical aperture (see Fig. \ref{fig:HIspec}b). 
We calculate the RMS spectral noise, in all the line-free channels in the velocity range, 3940 \kmps $<v< 6961$ \kmps, to be 6 and 9 mJy in the spectra with and without smoothing respectively.

We replicate the analysis in Fig. \ref{fig:HIspec} in the panels d and e of Fig. \ref{fig:BoxPlot}. 
We find that the profile width, $W_{20}$, is maximised and the velocity centroid of the profile remains coincident with the recessional velocity of ESO 137-G006. 
We replicate the analysis in Fig. \ref{fig:HItau} in panel f of Fig. \ref{fig:BoxPlot}.
The peak and integrated flux densities, absorption fractions and optical depths are all significant by more than $4 \sigma$ level, consistent with those reported in sections \ref{sec:HIfit} and \ref{sec:NHI}. 
There are no statistically significant differences in the other estimated properties, such as the column density, mass and spin temperature.
Overall, regardless of the chosen H{\sc i} detection region and application of Hanning smoothing, we find dense cold H{\sc i} gas in absorption against the core of Norma's BCG, ESO 137-G006.

\begin{table*}
\centering
\caption{Summary of the H{\sc i} properties measured from two different detection apertures, using ATCA spectra without and with Hanning smoothing.}
\label{tab:HIprop}
\begin{tabular}{|r|c|c|c|c|l|}

\textbf{Value} & \textbf{Ellipse} & \textbf{Ellipse + Smoothed} & \textbf{Square} & \textbf{Square + Smoothed} & \textbf{Unit} \\ \hline \hline
R.A. & 16:15:03.7 & 16:15:03.7 & 16:15:04 & 16:15:04 & J2000 \\ \hline
Dec & $-$60:54:24.9 & $-$60:54:24.9 & $-$60:54:26 & $-$60:54:26 & J2000 \\ \hline
Detection aperture & Ellipse at $31^{\circ}$:  $18 \times 14$ & Ellipse at $31^{\circ}$: $18 \times 14$  & Square: $22 \times 22$ & Square: $22 \times 22$ & \sqarcsec\\ \hline
Hanning kernel & $-$ & $3$ & $-$ & $3$ &  \\ \hline
Average spectral resolution ($\Delta v$) & $6.8$ & $6.8$ & $6.8$ & $6.8$ & \kmps \\ \hline
Starting profile edge ($v_{\rm{start}}$)& $5324 \pm 7$ & $5324 \pm 7$ & $5358 \pm 7$ & $5358 \pm 7$ & \kmps \\ \hline
Ending profile edge ($v_{\rm{end}}$) & $5556 \pm 7$ & $5556 \pm 7$ & $5557 \pm 7$ & $5557 \pm 7$ & \kmps \\ \hline
RMS spectral noise ($\sigma_{S}$) & $5$ & $3$ & $9$ & $6$ & mJy \\ \hline
Peak flux density ($S_{\rm{peak}}$) & $-31 \pm 5$ & $-28 \pm 3$ & $-48 \pm 9$ & $-41 \pm 6$ & mJy \\ \hline
Velocity at peak ($v_{\rm{peak}}$) & $5451 \pm 7$ & $5451 \pm 7$ & $5451 \pm 7$ & $5451 \pm 7$ & \kmps \\ \hline
Minimum width at 50\% of peak ($W_{50}$) & $21 \pm 26$ & $21 \pm 15$ & $21 \pm 27$ & $21 \pm 20$ & \kmps \\ \hline
Maximum width at 20\% of peak ($W_{20}$) & $226 \pm 39$ & $150 \pm 22$ & $185 \pm 40$ & $171 \pm 29$ & \kmps \\ \hline
Velocity centroid ($v_{\rm{HI}}$) & $5440 \pm 13$ & $5444 \pm 7$ & $5461 \pm 13$ & $5454 \pm 10$ & \kmps \\ \hline
Integrated flux density ($S_{\rm{HI}}$) & $-1.9 \pm 0.4$ & $-1.9 \pm 0.3$ & $-3.0 \pm 0.7$ & $-2.9 \pm 0.5$ & Jy \kmps \\ \hline
Peak absorption fraction ($(\Delta S/S_{c})_{\rm{peak}}$) & $0.91 \pm 0.46$ & $0.82 \pm 0.40$ & $0.71 \pm 0.35$ & $0.60 \pm 0.28$ &  \\ \hline
Total absorption fraction ($(\Delta S/S_{c})_{\rm{total}}$) & $0.25 \pm 0.03$ & $0.24 \pm 0.03$ & $0.23 \pm 0.03$ & $0.23 \pm 0.03$ &  \\ \hline
Peak optical depth ($\tau_{\rm{peak}}$) & $2.4 \pm 0.4$ & $1.7 \pm 0.3$ & $1.2 \pm 0.3$ & $0.9 \pm 0.2$ &  \\ \hline
Integrated optical depth ($\int\tau_{\rm{HI}}$) & $82 \pm 13$ & $74 \pm 8$ & $56 \pm 11$ & $54 \pm 7$ &  \\ \hline
Absorption column density ($N_{\rm{HI,abs}}$) & $1.5 \pm 0.2$ & $1.3 \pm 0.2$ & $1.0 \pm 0.2$ & $1.0 \pm 0.1$ & $10^{20} \, T_{\rm{spin}} \, \rm{cm}^{-2}$ \\ \hline
Non-detection limit ($3\sigma_{S}$) & $15$ & $10$ & $27$ & $18$ & mJy \\ \hline
Integrated emission flux ($S_{\rm{int}}$) & $< 2.3$ & $< 1.5$ & $< 4.0$ & $< 2.6$ & Jy \kmps \\ \hline
H{\sc i} mass ($M_{\rm{HI}}$) & $< 2.7$ & $< 1.7$ & $< 4.7$ & $< 3.0$ & $10^{9} \, \mathrm{M}_{\odot}$ \\ \hline
Brightness temperature ($T_{\rm{B}}$) &  $< 2.2$ &  $< 1.4$ &  $< 3.9$ &  $< 2.5$ & $10^{4}$ K \\ \hline

Emission column density ($N_{\rm{HI,em}}$) & $< 4.1$ & $< 2.6$ & $< 7.1$ & $< 4.6$ & $10^{22} \, \rm{cm}^{-2}$ \\ \hline
Spin temperature ($T_{\rm{spin}}$) & $< 273$ & $< 194$ & $< 689$ & $< 471$ & K \\ \hline
\end{tabular}
\end{table*}

\section{Comparison to HIPASS Observations}
\label{sec:HIPASS}

\begin{figure*}
    \centering
    \includegraphics[width=\textwidth]{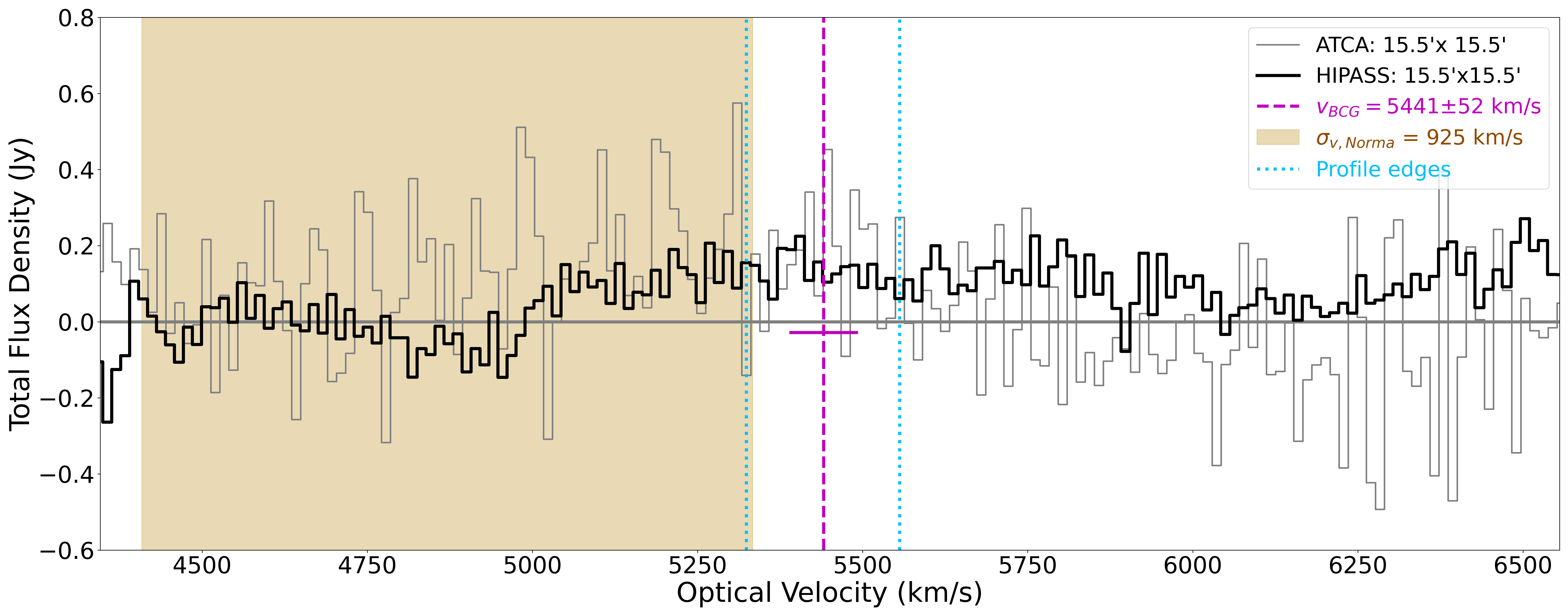}
    \caption{Spectra from ATCA and HIPASS spectral cubes integrated over an aperture of the size of one HIPASS synthesised beam (15.5 arcmin) centred on ESO 137-G006's radio core.
    The edges of the detected H{\sc i} absorption profile, ESO 137-G006's recessional velocity and the Norma cluster's velocity dispersion are marked as in Fig. \ref{fig:HIspec}. 
    This figure shows that it is not possible to resolve the detected H{\sc i} absorption profile towards ESO 137-G006's core using HIPASS data.}
    \label{fig:HIPASSspec}
\end{figure*}

The H{\sc i} absorption detection made in this study, using ATCA observations towards ESO 137-G006's core, was not detected using the HIPASS data. 
To verify this, we extract a `HIPASS spectrum' from the HIPASS spectral cube, over a single circular HIPASS synthesised beam centred on ESO 137-G006's radio core (FWHM 15.5 arcmin, dashed-brown circle in Fig. \ref{fig:Maps}b).
Fig. \ref{fig:HIPASSspec} presents this HIPASS spectrum in thick black.
To compare, we also extract an `ATCA spectrum' from the ATCA spectral cube over the same region, \ie a circle of diameter 15.5 arcmin centred on R.A. (J2000) = 16:15:03.7 and Dec (J2000) = $-60$:54:24.9.
Since the HIPASS spectrum has an optical velocity resolution of $\approx 14$ \kmps, we bin the ATCA spectrum, with bin-width = 2, to match the spectral resolution.
Fig. \ref{fig:HIPASSspec} presents this binned ATCA spectrum in thin grey.
No Hanning smoothing or baseline subtraction are applied on either spectra.
In Fig. \ref{fig:HIPASSspec}, we display the same velocity range as Fig. \ref{fig:HIspec}a to compare the velocities of the host galaxy and cluster with that of the detected profile.
The detected profile's edges (vertical cyan dotted lines), ESO 137-G006's recessional velocity (vertical magenta dashed line and error-bar) and Norma cluster's velocity dispersion (vertical golden shaded region) are marked for comparison.
We find that the H{\sc i} absorption profile, detected using the ATCA data, is smeared out by the more than 100 times larger HIPASS synthesised beam. 
The Parkes beam encompasses all of ESO 137-G006's radio continuum structure and cannot resolve the location of H{\sc i} against the much smaller radio core (see Fig. \ref{fig:Maps}b). 
With this comparison, we conclude that it is not possible to obtain a firm detection of H{\sc i} absorption towards ESO 137-G006's core based on HIPASS alone.

\bsp	
\label{lastpage}
\end{document}